\documentclass[reprint,aps,prb,amsmath,amssymb,longbibliography, notitlepage]{revtex4-1}
\usepackage{amsmath,amssymb, bm, graphicx}
\usepackage{float} 
\usepackage{algorithm}
\usepackage[stable]{footmisc}
\usepackage{soul}
\usepackage{physics}
\usepackage{pifont}
\usepackage[normalem]{ulem}
\usepackage{gensymb}
\usepackage{multirow}
\usepackage{tabularx}
\usepackage[colorlinks, linkcolor= blue, citecolor = blue, urlcolor=blue]{hyperref}
\usepackage{array}
\usepackage[normalem]{ulem}

\newcolumntype{P}[1]{>{\centering\arraybackslash}p{#1}}
{
\graphicspath{ {./figures/} }
\def\nn{\nonumber}
\def\bea{\begin{eqnarray}}
\def\eea{\end{eqnarray}}
\def\be{\begin{equation}}
\def\ee{\end{equation}}

\def\kb{{\bm k}}

\def\e{\varepsilon}
\def\m{\mathcal}
\def\bal{\begin{aligned}}
\def\eal{\end{aligned}}

\begin{document}
\title{Intrinsic nonlinear Nernst and Seebeck effect}
\author{Harsh Varshney}
\email{hvarshny@iitk.ac.in}
\affiliation{Department of Physics, Indian Institute of Technology, Kanpur-208016, India.}
\author{Amit Agarwal}
\email{amitag@iitk.ac.in}
\affiliation{Department of Physics, Indian Institute of Technology, Kanpur-208016, India.}
% 
%%%%%%%%%%%%%%%%%%%%%%%%%%%
\begin{abstract}  

The Nernst and Seebeck effects are crucial for thermoelectric energy harvesting. However, the linear anomalous Nernst effect requires magnetic materials with intrinsically broken time-reversal symmetry. In non-magnetic systems, the dominant transverse thermoelectric response is the nonlinear Nernst current. Here, we investigate nonlinear Nernst and Seebeck effects to reveal intrinsic scattering-free Seebeck and Nernst currents arising from band geometric effects in bipartite antiferromagnets (parity-time-reversal symmetric systems). We show that these contributions, independent of scattering time, originate from the Berry connection polarizability tensor which depends on the quantum metric. Using ${\rm CuMnAs}$ as a model system, we demonstrate the dominance of intrinsic nonlinear Seebeck and Nernst currents over other scattering-dependent contributions. Our findings deepen the fundamental understanding of nonlinear thermoelectric phenomena and provide the foundation for using them to develop more efficient, next-generation energy harvesting devices.
\end{abstract}

%%%%%%%%%%%%%%%%%%%%%%%%%%%
\maketitle

\section{Introduction}

Temperature gradients and heat flow are omnipresent in all systems, presenting significant opportunities for generating electrical energy through thermoelectric effects. By converting heat, often from waste sources, into electricity, thermoelectric energy harvesting holds significant promise for sustainable technology development~\cite{Snyder2009thermoelectric, enescu2019green, amin2020review,tan2011sustainable, massetti2021unconventional, koumoto2013thermoelectric, sothmann2014thermoelectric,chhatrasal2016, takao2021, syed2024, shenmultifunctional, Yu2024ambient}. The Seebeck and Nernst effects are central to this effort. The transverse Nernst and longitudinal Seebeck effects generate electric currents and voltage in response to an applied temperature gradient. Typically, the Nernst effect requires an external magnetic field. In the absence of a magnetic field, the anomalous Nernst effect can arise in magnetic systems~\cite{kamran2016nernst, checklesky2009thermpower, ikhlas2017large, lee2014seebeck, kamal2020thermal, natalya2016seebeck, uchida2008observation, Rana2018thermopower, kamal2019berry, sunit2023chiral, win2024thermoelectric, theja2024,kurschner2024thermoelectric, Pasquale2024, Weinan_Seebeck, Sakai_Nature}. 
% \textcolor{red}{Please cite Kamal's two more papers on Chiral anomaly here}. 
However, realizing this effect remains challenging in non-magnetic (time-reversal symmetric) systems, where the linear anomalous Nernst effect vanishes. 

\begin{figure}
    \centering
    \includegraphics[width = \linewidth]{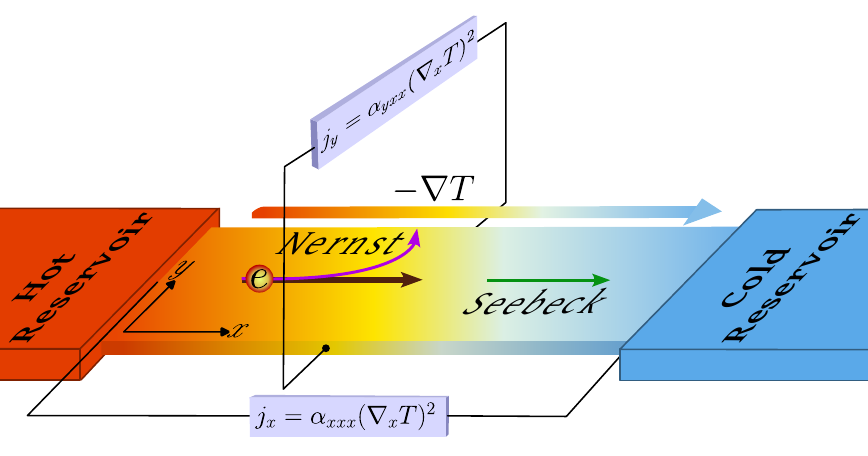}
    \caption{{\bf Nonlinear Nernst and Seebeck effect $\propto (\nabla T)^2$.} Schematic depicting nonlinear thermoelectric response in a two-dimensional system subjected to a temperature gradient. The temperature gradient can be induced via optical, electrical, or thermal means.}
    \label{fig:schematics}
\end{figure}

This challenge has led to growing interest in the nonlinear Nernst current~\cite{Arisawa2024observation}, which dominates the thermoelectric response in non-magnetic systems without a magnetic field [see Fig.~\ref{fig:schematics}]. Recent studies have predicted the existence of a Berry curvature dipole-induced nonlinear Nernst effect in non-magnetic systems with broken inversion symmetry, though the nonlinear Seebeck current is absent in these systems~\cite{zeng2019nonlinear, yu2019topological, karki2019nonlinear}. This raises the question: Are there other band-geometry-induced contributions to the nonlinear Nernst and Seebeck effects? Recent works on nonlinear electric currents in response to an applied electric field have revealed an intrinsic contribution, independent of scattering time, originating from the quantum metric — a geometric property of electronic wavefunctions~\cite{kamal2022resonant, kamal2023intrinsic, wnag2021intrinsic, wang2024intrinsic, anyuan2023quantum, wang2023quantum, liu2021intrinsic, Sinha2022berry, Adak2024tunable, Datta2024nonlinear}.  
% \textcolor{red}{Please cite our N. Phys, and N. Mat review paper here}. 
Interestingly, this contribution dominates the charge response in bipartite antiferromagnets, where parity-time-reversal symmetry is preserved.

In this paper, we investigate the nonlinear Nernst and Seebeck effect to demonstrate the existence of quantum metric-induced scattering time independent {\it intrinsic} nonlinear Nernst and Seebeck currents. We present a comprehensive analysis of all contributions to the second-order nonlinear Nernst and Seebeck effect, highlighting their dissipative and dissipationless components. We show that in parity-time-reversal symmetric systems like CuMnAs, the intrinsic contributions to nonlinear Nernst and Seebeck conductivities can dominate thermoelectric transport. Our work opens new avenues for waste energy harvesting through nonlinear thermoelectric effects.

%%%%%%%%%%%%%%%%%%%%%%%%%%%%%%
%%%%%%%%%%%%%%%%%%%%%%%%%%%%%%

%%%%%%%%%%%%%%%%%%%%%%%%%%%%%%%%%%
%%% |p{2cm}|p{7cm}|p{9cm}|
%%{\linewidth}{| >{\centering\arraybackslash}X | >{\centering\arraybackslash}X | >{\centering\arraybackslash}X |}

%%%%%%%%%%%%%%%%%%% making a table  %%%%%%%%%%%
\begin{table*}[t!]
\caption{We present the three distinct contributions to the nonlinear thermoelectric conductivity [see Eq.~\eqref{eq:nl_cond}], expressed in terms of the dissipative (Ohmic) and dissipationless (Hall) parts. $\bar{\alpha}_{a;bc}$ denotes the second-order thermoelectric conductivity tensor symmetrized in the last two field indices $b$ and $c$. The dissipative or Ohmic (dissipationless or Hall) contribution to the nonlinear thermoelectric conductivity tensor is marked by the superscript O (H). The Drude contribution originates from the band velocity and varies quadratically with the scattering timescale. The Berry curvature contribution is proportional to $\tau$. The intrinsic contribution is $\tau-$independent and it arises from the Berry connection polarizability which depends on the quantum metric of the system.}
%denotes the Ohmic or dissipative (Hall or dissipationless) component of .}

\centering
\begin{tabular}{P{2cm}  P{8 cm} P{8cm}}
\hline\hline
     \multirow{2}{5em}{ \centering Origin} & \multicolumn{2}{c}{Nonlinear thermoelectric conductivity (${\bar \alpha}_{a;bc}$) in units of $e/\hbar T^2$} \\[1ex] \cline{2-3}
     &  Dissipative part $({\bar \alpha}^{\rm O}_{a;bc})$  & Dissipationless part $({\bar \alpha}^{\rm H}_{a;bc})$ \\ \hline \hline
    {\centering Drude (band velocity)} & $-\frac{  \tau^2}{2 \hbar^2} \sum_{n, \kb}  \tilde\e_n  \partial_a \e_n \partial_b \e_n \partial_c \e_n  \left( \partial_{\e_n} f_n + \tilde\e_n \partial^2_{\e_n} f_n \right)$ & 0\\ \hline
     {\centering Anomalous (Berry curvature)}& {\centering $0$} & $-\frac{\tau}{2\hbar} \sum_{n, \kb} \tilde\e_n^2 \partial_{\e_n} f_n  (\Omega^{ab}_n \partial_c\e_n + \Omega^{ac}_n \partial_b\e_n)$ \\ \hline
     \multirow{2}{6em}{ \centering Intrinsic (Berry connection polarizability)} & $\frac{1}{2} \sum_{n,p,\kb}^{n \ne p} \tilde\e_n(\tilde\e_n + \tilde\e_p)f_n (\partial_a{\tilde{\m G}}^{bc}_{np} + \partial_b{\tilde{\m G}}^{ac}_{np} + \partial_c{\tilde{\m G}}^{ab}_{np}) $ & $- \frac{1}{2}\sum_{n,p,\kb}^{n \ne p} \tilde\e_n(\tilde\e_n + \tilde\e_p)f_n (2\partial_a{\tilde{\m G}}^{bc}_{np} - \partial_b{\tilde{\m G}}^{ac}_{np} - \partial_c{\tilde{\m G}}^{ab}_{np} )$ \\ 
      & ~~~~~$ + \frac{1}{6}\sum_{n, p, \kb}^{n \ne p} \tilde\e_n f_n [ \{2\partial_a{\m G}^{bc}_{np} + {\tilde{\m G}}^{bc}_{np}(\partial_a\e_p + 5\partial_a\e_n)\}  + a \leftrightarrow b + a \leftrightarrow c ]$ &~~~~~$-\frac{1}{3}\sum_{n, p, \kb}^{n \ne p} \tilde\e_n f_n [ \{ 2\partial_a{\m G}^{bc}_{np} + {\tilde{\m G}}^{bc}_{np}(\partial_a\e_p + 5\partial_a\e_n)\} - \frac{1}{2}\{ a \leftrightarrow b \} - \frac{1}{2}\{a \leftrightarrow c\} ]$ \\  \hline \hline
\end{tabular}
\label{tab:table1}
\end{table*}
%%%%%%%%%%%%%%%%%%%%%%%%%%%%%%%%%%
\section{Nonlinear thermoelectric current}
We use the quantum kinetic theory framework based on the density matrix to capture the role of the band geometric quantities such as the Berry curvature and quantum metric in thermoelectric responses. 
%Before proceeding with this, we briefly discuss the subtle issue of the physically observable current in the semiclassical description of electrical currents generated by a temperature gradient. 
In the semiclassical theory of charge transport, the charge carriers or quasi-particles are wave packets that translate and rotate in the phase space. The translational motion generates a transport current, while the rotational motion generates a circulating current, with their sum being the net local current. However, in experiments, the circulating current (also known as the magnetization current) is not observable and only the transport current is measured. In the presence of a temperature gradient, the local and magnetization electrical current densities~\cite{xiao2006berry, kamal2021intrinsic} are given as ${\bm j}^{\rm loc} = -e \int [d\kb] \dot{{\bm r}} f_\kb +  {\bm \nabla}_r \times \int[d\kb] {\bm m} f_\kb $ and ${\bm j}^{\rm mag} = {\bm \nabla}_r \times {\bm M}$. Here, $-e$ ($e>0$) is the charge of the electron, $\dot{\bm r}$ is the velocity of the center of mass of the wavepacket, $f_\kb$ represents the nonequilibrium distribution function, ${\bm m}$ denotes the orbital magnetic moment, and ${\bm M}$ is the orbital magnetization density. Additionally, $[d\kb] \equiv d^d\kb/(2\pi)^d$ denotes the integration measure of the $d$-dimensional space.  The orbital magnetization density is specified by~\cite{xiao2006berry, kamal2021intrinsic, rhonald2024orbital} ${\bm M} = \int [d\kb] \left( \pdv{g_1}{\e_\kb} ~ {\bm m} - \frac{e}{\hbar} g_1 {\bm \Omega} \right)$, where $\bm \Omega$ is the Berry curvature, $g_1 \equiv -k_B T \log[ 1 + e^{-\beta(\e_\kb - \mu)}]$ is the grand potential density of a particular state, and $\pdv{g_1}{\e_\kb} = f_\kb$~. Note that all the physical quantities in the expressions above are for a specific band, and we have suppressed the band index for brevity. 
The observable electrical current is obtained by subtracting the magnetization current density from local current density, ${\bm j} = {\bm j}^{\rm loc} - {\bm j}^{\rm mag}$, to yield, ${\bm j} = - e \int[d\kb] \dot{{\bm r}} f_\kb  + \frac{e}{\hbar} {\bm \nabla}_r \times \int[d\kb] g_1 {\bm \Omega}~$. 

We follow a similar approach to the quantum kinetic theory. Within the quantum kinetic theory, the electrical current generated by an applied temperature gradient ~\cite{sekine2020quantum} is given by, 
\bea\label{eq:anc-def}
{\bm j} &=& -e {\rm Tr}[\hat{\bm v} \rho] +  \sum_n {\bm E}_T \times {\bm M}^n_{\Omega}~.
%
% & = & - e \sum_{n,p,\kb} \hat{\bm v}_{pn} \rho_{np} + {\bm E}_T \times {\bm M}_{\Omega}~.
\eea 
Here, Tr denotes the trace of a matrix, $\hat{\bm v} = (i/\hbar)[{\m H}_0, \hat{\bm r}]$ is the velocity operator with ${\cal H}_0$ being the unperturbed Hamiltonian of the system and $\hat{\bm r}$ denotes the position operator, and $\rho$ represents the density matrix. We express the temperature gradient in terms of a thermal field specified by ${\bm E}_T \equiv - {{\bm \nabla} T}/T$ and ${\bm M}^n_\Omega$ is the Berry curvature-induced contribution to the orbital magnetization density of the $n$th energy band of ${\m H}_0$. This quantum kinetic generalization of the semiclassical theory reproduces all the known results in the linear response regime. More interestingly, it also accounts for multi-band coherence effects and has been recently shown to give rise to novel intrinsic contributions to thermal transport. 

%has been shown to 
%The above equation resembles the semiclassical description. the only difference is that the distribution function $f_\kb$ is replaced by density matrix and the del operator (${\bm \nabla}_r$) by $- {\bm E}_T$. 
%Thus, we can consider the quantum kinetic theory framework as a generalization of semiclassical theory, which accounts for interband or coherence effects. 

To calculate all contributions to the thermoelectric currents up to second-order in the applied temperature gradient, we compute the density matrix through the quantum-Liouville equation~\cite{harsh2023quantum, harsh2023intrinsic}, $\pdv{\rho(\kb,t)}{t} + \frac{i}{\hbar} [{\cal H}, \rho(\kb,t)] = 0$. In this equation, $\rho(\kb, t)$ is the density matrix which depends on the crystal momentum $\kb$ and time $t$. We will denote it as $\rho$ for brevity. 
Furthermore, ${\m H} = {\m H}_0 + {\m H}_T + {\m H}_{\rm dis}$ is the total Hamiltonian of the system.  Here, ${\m H}_T$ and ${\m H}_{\rm dis}$ are the temperature gradient and disorder contribution to the unperturbed Hamiltonian ${\m H}_0$, whose eigenvalues and eigenvectors are known. These eigenvectors of ${\m H}_0$ form the basis set used for all our calculations of the density matrix.  
% \textcolor{red}{It can be decomposed into three parts as ${\m H} = {\m H}_0 + {\m H}_T + {\m H}_{\rm dis}$. Here, ${\m H}_T$ and ${\m H}_{\rm dis}$ are the temperature gradient and disorder contribution to the bare Hamiltonian ${\m H}_0$, whose eigenvalues and eigenvectors are known.}

We include the effect of the temperature gradient through perturbation theory and treat the disorder part of the Hamiltonian within the relaxation time approximation. To obtain the density matrix elements up to second-order in the temperature gradient, we express the density matrix as $\rho = \sum_{N = 0}^{\infty} \rho^{(N)}$ where $\rho^{(N)} \propto (\nabla T)^N$ represents the $N$th order contribution. 
For instance, $\rho^{(0)}$ represents the equilibrium density matrix independent of the temperature gradient, while $\rho^{(1)}$ and $\rho^{(2)}$ denote first and second-order density matrix contributions, respectively. 
Using this, we obtain an iterative quantum-Liouville equation to calculate the $N$th-order density matrix  element~\cite{harsh2023intrinsic, harsh2023quantum},   
\be\label{eq:dme} 
\pdv{\rho^{(N)}_{np}}{t} + \frac{i}{\hbar}[{\m H}_0, \rho^{(N)}]_{np} + \frac{\rho^{(N)}_{np}}{\tau/N} = [D_T(\rho^{(N-1)})]_{np}~.
\ee 
Here, $\tau$ captures the scattering time of the Bloch electrons, and for simplicity we consider it to be a constant. 
The subscripts denote the band indices for the Hamiltonian, satisfying ${\m H}_0 \ket{u_n(\kb)} = \tilde\e_n \ket{u_n(\kb)}$, where $\tilde\e_n = (\e_n(\kb) - \mu)$ with $\e_n(\kb)$ being the energy of the $n$th energy band and $\mu$ denotes the chemical potential. Here, $\ket{u_n(\kb)}$ is the cell periodic part of the Bloch wavefunction of the $n$th energy eigenstate. In Eq. ~\eqref{eq:dme}, $D_T(\rho)$ represents the thermal driving term defined as~\cite{sekine2020quantum, harsh2023intrinsic, harsh2023quantum}  
\begin{equation}
D_T(\rho) = -\frac{1}{2\hbar}{\bm E}_T \cdot \left[\{ {\m H}_0, \partial_\kb \rho\} - i [{\bm {\m R}}_\kb, \{ {\m H}_0, \rho\}]\right]~.
\end{equation}
Here, ${\bm {\m R}}_\kb$ is the momentum space Berry connection, and the brackets $[\cdot, \cdot]$ and $\{\cdot, \cdot\}$ denote commutator and anti-commutator, respectively.

%%%%%%%%%%%%%%%%%%%%%%%%%%%%%%
\begin{figure}[t]
    \centering
    \includegraphics[width = \linewidth]{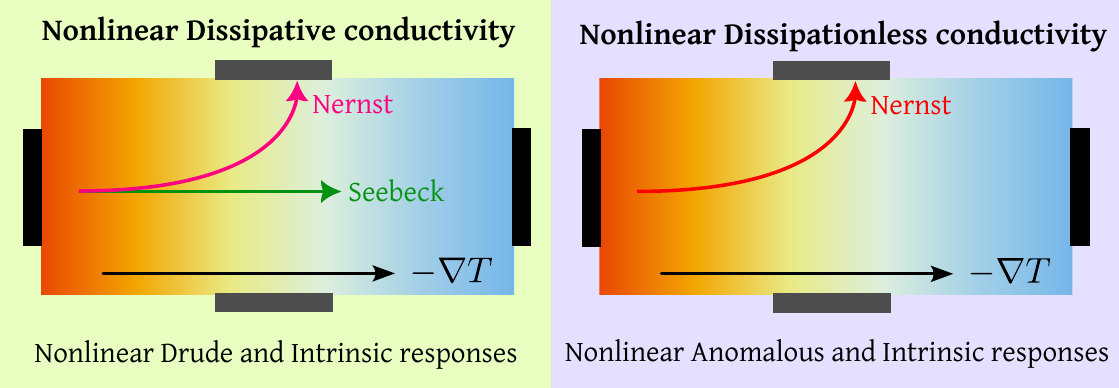}
    \caption{{\bf The nonlinear dissipative and dissipationless conductivities}. The orange-blue color gradient indicates the temperature gradient, with orange being the hotter side, which induces a thermoelectric charge current. The Seebeck effect is always dissipative, and it can be induced by either the Drude or the intrinsic Berry connection polarizability contribution. The Drude contribution to the nonlinear Nernst effect is always dissipative, while the Berry curvature-induced contribution is always dissipationless. In contrast, the Berry connection polarizability can induce both dissipative and dissipationless nonlinear Nernst effect. 
    %The black and gray strips depict the contacts made for probing nonlinear Seebeck and Nernst currents, respectively.  }
    % \textcolor{red}{The figure needs to be checked again and changed, along with the table.}
    \label{fig:HO_schematic}}
\end{figure}
%%%%%%%%%%%%%%%%%%%%%%%%%%%%%%

%After establishing a framework for calculating $N$th order density matrix iteratively through 
We use Eq.~\eqref{eq:dme} to compute first and second-order density matrix elements in the steady state where the time derivative of the density matrix can be ignored. 
The matrix elements of the commutator relation ${\m H}_0$ and $\rho^{N}$ are given by $ [{\m H}_0, \rho^{(N)}]_{np} = \e_{np} \rho^{(N)}_{np}$, where $ \e_{np} = (\e_n -\e_p)$ is the energy difference between the bands at a given $\kb$. Using this and Eq.~\eqref{eq:dme}, we obtain the following iterative equation for the $N$th order density matrix,  
\be\label{eq:Nth_dm}
\rho^{(N)}_{np} = -i\hbar g_N^{np} [D_T(\rho^{(N-1)})]_{np}~.
\ee
Here, we have defined $g_{N}^{np} \equiv (\e_{np} -i\hbar N/\tau )^{-1}$, which carries the information of the scattering time. We present the details of the calculations of the first and second-order density matrix in Sec.~S1 of the Supplementary Material (SM)~\footnote{The Supplementary Material discusses: S1) the derivation of the first and second-order density matrix in the presence of a temperature gradient, S2) calculation of the linear and nonlinear thermoelectric currents, S3) heat dissipation in thermoelectric processes, S4) the dissipative and dissipationless parts of the nonlinear thermoelectric current.}. We decompose the first-order density matrix into diagonal and off-diagonal parts as $\rho^{(1)} = \rho^{\rm d} + \rho^{\rm o}$. As each of these terms contributes to both diagonal and off-diagonal parts of the second-order density matrix, we express the second-order density matrix as $\rho^{(2)} = \rho^{\rm dd} + \rho^{\rm do} + \rho^{\rm od} + \rho^{\rm oo}$. Here, the first two terms are the diagonal parts of the second-order density matrix, while the last two terms denote the off-diagonal parts. In this decomposition, the first superscript index represents the diagonal or off-diagonal part of $\rho^{(2)}$. The second superscript index denotes the diagonal or off-diagonal part of $\rho^{(1)}$ which gives rise to the respective component of the second-order density matrix. 

Two of these four contributions to the second-order density matrix are given by, 
%$\rho = \rho^{dd} + \rho^{do} +\rho^{od} + \rho^{oo} $. We have, 
%
\bea 
\rho^{\rm{dd}}_{nn} &=&  \frac{\tau^2}{2\hbar^2} \left[ \hbar \tilde\e_n v^n_{b} \partial_c f_n  +  \tilde\e^2_n \partial_b \partial_c f_n \right] E^b_T E^c_T~,  \\ 
\rho^{\rm{od}}_{np} &=& \frac{\tau}{\hbar}g^{np}_2 \m{R}_{np}^{b}\left(\tilde{\e}_n ^2 \partial_{c}f_{n}-\tilde{\e}_p^2 \partial_{c}f_{p} \right)  E^b_T E^c_T~. 
\eea 
Here, $v^b_n = \hbar^{-1} \partial_b \e_n$ represents the band velocity component of Bloch electrons along the $b$ direction with $\partial_b \equiv \pdv{}{\kb_b}$. ${\m R}^b_{np} = i \bra{u_n}\partial_b \ket{u_p}$ is the momentum-space interband Berry connection. 
The other two components of $\rho^{(2)}$ are given by, 
\begin{widetext}
\bea 
\rho^{\rm{do}}_{nn} &=& \frac{i\tau}{4\hbar^2}\sum_{p}^{p \ne n} (\tilde{\e}_n + \tilde{\e}_p) \left(g^{np}_1\m{R}_{np}^{c}\m{R}_{pn}^{b} + g^{pn}_1\m{R}_{np}^{b}\m{R}_{pn}^{c}\right)  \xi_{np} E^b_T E^c_T~,  \\ 
\rho^{\rm{oo}}_{np}  &=& 
\frac{-i}{2}g^{np}_2 (\tilde{\e}_n + \tilde{\e}_p ) \m{D}^{b}_{np}\left(g^{np}_1\m{R}_{np}^{c}\xi_{np}\right) E^b_T E^c_T + \frac{1}{ 2}g_2^{np} \sum_{q\neq n \neq p} \left[ g^{nq}_1\m{R}^{c}_{nq}\m{R}^{b}_{qp}(\tilde{\e}_n +\tilde{\e}_q )\xi_{nq} -  g^{qp}_1\m{R}^{b}_{nq}\m{R}^{c}_{qp}(\tilde{\e}_q +  \tilde{\e}_p )\xi_{qp}\right]E^b_T E^c_T~. \nn \\ 
\eea 
\end{widetext}
Here, we have defined $\xi_{np} = \tilde\e_n f_n - \tilde\e_p f_p$, where $f_n = (1 + e^{\beta \tilde\e_n})^{-1}$ is the Fermi-Dirac distribution function with $\beta = 1/(k_B T)$ being the inverse-temperature. We defined ${\m D}^b_{np} = \partial_b - i ({\m R}^b_{nn} - {\m R}^b_{pp})$ as the covariant derivative.

We use the obtained $\rho^{(2)}$ to calculate the linear and nonlinear anomalous thermoelectric conductivity specified by Eq.~\eqref{eq:anc-def}. We recast Eq.~\eqref{eq:anc-def} as $ {\bm j} =  - e \sum_{n,p,\kb} \hat{\bm v}_{pn} \rho_{np} + \sum_n {\bm E}_T \times {\bm M}^n_{\Omega}~.$ In the Bloch band basis, the velocity operator's elements satisfy $\hat{\bm v}_{pn} = \hbar^{-1} ( \partial_\kb \e_p \delta_{pn} + i \bm{\m R}_{pn} \e_{pn})$ with $\delta_{pn}$ being the Kronecker delta function. We present the detailed calculation in Sec.~S2 of the SM~\cite{Note1}. 

%%%%%%%%%%%%%%%%%%%%%%%%%%%%%%%%%%%%%%%%%%%%%%%%%%
\section{Intrinsic thermoelectric conductivity}

The nonlinear electrical current can be expressed in terms of the thermal field, via the relation, $j^{(2)}_a = T^2\alpha_{a;bc} E_T^b E_T^c = \alpha_{a;bc} \nabla_b T ~\nabla_c T$. 
Using this relation, we extract the nonlinear thermoelectric conductivity tensor of third rank, $\alpha_{a;bc}$. The first index in $\alpha_{a;bc}$ denotes the current direction, while the last two indices after the semicolon denote the thermal field indices. 
The general expression of the total thermoelectric response obtained by us is presented in Sec.~S2 of the SM \cite{Note1}. 
In the clean limit where the scattering time corresponds to the smallest energy scale, such that $\tau \e_{np}/\hbar \gg 1$, we can simplify the obtained nonlinear thermoelectric conductivity in three distinct contributions. 
We obtain, $\alpha_{a;bc} = \alpha_{a;bc}^{\rm NLD} + \alpha_{a;bc}^{\rm NLA} + \alpha_{a;bc}^{\rm NLI}$, where each contribution has a distinct $\tau$ dependence. The nonlinear Drude ($\alpha^{\rm NLD}$) conductivity varies quadratically with $\tau$, the nonlinear anomalous ($\alpha^{\rm NLA}$) conductivity scales linearly with $\tau$, while the nonlinear intrinsic ($\alpha^{\rm NLI}$) conductivity is independent of $\tau$. The intrinsic thermoelectric conductivity is given by, 
\bea\label{eq:nl_cond} 
\alpha^{\rm NLI}_{a;bc} &=& \frac{e}{2\hbar T^2} \sum_{n, p, \kb}^{n \ne p}  \tilde\e_n f_n \Big[ -(\tilde\e_n + \tilde\e_p) \left( \partial_a{\tilde{\m G}^{bc}_{np}} - 4\partial_c{\tilde{\m G}^{ab}_{np}}\right) \nn \\ 
&& + 2 \left(  2\partial_c{\m G}^{ab}_{np} + \tilde{\m G}^{ab}_{np} (\partial_c \e_p + 5\partial_c \e_n )  \right) \Big]~.
\eea 
Here, $\tilde{\m G}^{ab}_{np} \equiv {\m G}^{ab}_{np}/\e_{np}$ is related to the band-normalized quantum metric, and it is also known as the Berry connection polarizability (BCP) tensor~\cite{liu2021intrinsic}. The band-resolved quantum metric tensor~\cite{watanabe2021chiral} is specified by ${\m G}^{ab}_{np} = ({\m R}^{a}_{np} {\m R}^{b}_{pn} + {\m R}^{a}_{pn} {\m R}^{b}_{np})/2$, which is symmetric under the exchange of both spatial and band indices, $i.e.,$ ${\m G}^{ab}_{np} = {\m G}^{ba}_{np}$ and ${\m G}^{ab}_{np} = {\m G}^{ab}_{pn}$. This scattering independent contribution has not been explored earlier and it is one of the main results of this manuscript. 
We show below that this Berry connection polarizability-induced contribution will give rise to intrinsic nonlinear Nernst and Seebeck effect. 
%The quantity $\tilde{\m G}^{ab}_{np} \equiv {\m G}^{ab}_{np}/\e_{np}$ is the band-normalized band-resolved quantum metric, which is also attributed as Berry connection polarizability (BCP) tensor~\cite{liu2021intrinsic}.

The nonlinear Drude and nonlinear anomalous contributions are given by, 
\begin{align}
\alpha^{\rm NLD}_{a;bc} &=-\frac{e \tau^2}{2 \hbar^3 T^2} \sum_{n, \kb}  \tilde\e_n \partial_a\e_n \partial_b\e_n \partial_c\e_n \pdv{~}{\e_n}\left( {\tilde \e_n} \pdv{f_n}{\e_n}\right)~, \\ 
\alpha^{\rm NLA}_{a;bc} &= -\frac{e\tau}{\hbar^2 T^2} \sum_{n,\kb} \tilde\e^2_n \Omega^{ab}_n \partial_c f_n~.
\end{align}
Here, $\Omega^{ab}_n$ is the Berry curvature of the $n$th band.  Note that the band-resolved Berry curvature~\cite{watanabe2021chiral} is defined as $\Omega^{ab}_{np} = i ({\m R}^{a}_{np} {\m R}^{b}_{pn} - {\m R}^{a}_{pn} {\m R}^{b}_{np})$, which is related to the single band Berry curvature by the relation, $\Omega_{n}^{ab} = \sum_{p}^{p \ne n} \Omega^{ab}_{np}$. The band-resolved Berry curvature is an anti-symmetric tensor in both spatial and band indices, $, i.e.,$ $\Omega^{ab}_{np} = -\Omega^{ba}_{np}$ and $\Omega^{ab}_{np} = -\Omega^{ab}_{pn}$, respectively. As a consistency check, we highlight that the expression of the nonlinear Drude and anomalous conductivities obtained by us is identical to that derived within the semiclassical framework in other studies~\cite{yu2019topological, zeng2019nonlinear, atasi2022nonlinear, wang2022quantum}. 
%Furthermore, the nonlinear intrinsic conductivity has a purely quantum mechanical origin and arises from the quantum metric tensor. 
We emphasize that the nonlinear Drude and nonlinear anomalous conductivities are the Fermi surface contributions as they depend on the derivatives of the Fermi function. In contrast, the nonlinear intrinsic contribution is a Fermi sea contribution and it depends only on the Fermi function, not its derivatives. Thus, the nonlinear intrinsic contribution can be finite within the bandgap of a material, where all other contributions vanish. 
%%%%%%%%%%%%% inserting figure %%%%%%%%%%%%%%%%%
\begin{figure}[t!]
    \centering
    \includegraphics[width = \linewidth]{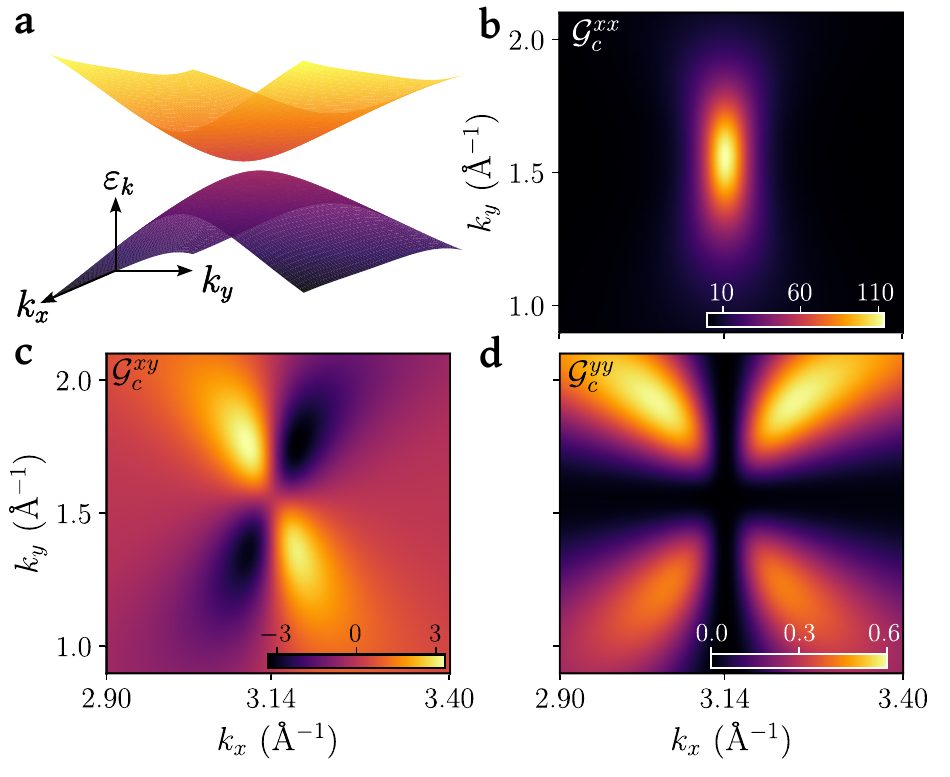}
    \caption{{\bf The quantum metric in CuMnAs}. \textbf{a.} The electronic band dispersion of ${\m PT}$ symmetric ${\rm CuMnAs}$, based on the Hamiltonian  in Eq.~\eqref{eq:ham}. %is plotted in units of eV. 
    \textbf{b-d.} The momentum space distribution of the different components of the quantum metric for the conduction band, in units of ${\rm \AA}^2$. We have used the following Hamiltonian parameters: $t = 0.08~$eV, and $\tilde{t} = 1~$eV along with  $\alpha_R = 0.8$, $\alpha_D = 0$, and ${\bm h}_{\rm AFM} = ( 0.85,0,0)~$eV. } 
    \label{fig:cumnas}
\end{figure}
%%%%%%%%%%%%%%%%%%%%%%%%%%%%%%%%%%%%%%%%%%%%%%%%%%%

%%%%%%%%%%%%%%%%%%%%%%%%%%%%%%
\begin{figure*}[t!]
    \centering
    \includegraphics[width = \linewidth]{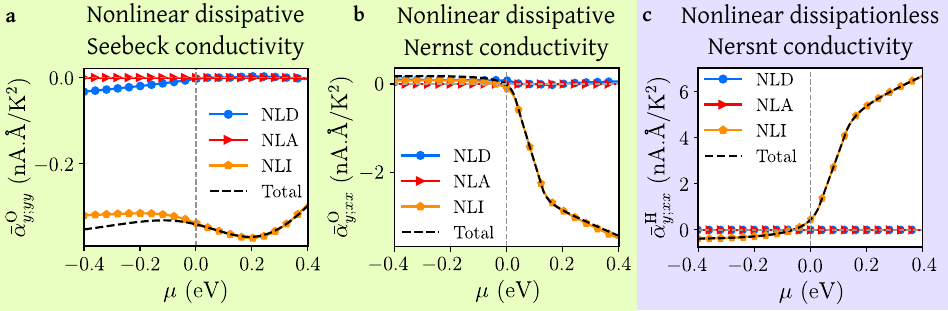}
    \caption{The variation of the nonlinear thermoelectric conductivity for ${\m PT}$ symmetric ${\rm CuMnAs}$ with the chemical potential ($\mu$).  The dissipative part of the nonlinear (\textbf{a}) Seebeck (longitudinal) conductivity, and (\textbf{b}) the Nernst conductivity. (\textbf{c}) The variation of the dissipationless nonlinear Nernst conductivity with the chemical potential. Here, the Hamiltonian parameters are the same as in Fig.~\ref{fig:cumnas}. Additionally, we have set the temperature $T = 50~K$, and chosen $\tau = 10^{-14}$s. 
    %{Here, the grey dashed lines are guides to the eye.}
    \label{fig:nl_cond}}
\end{figure*}
%%%%%%%%%%%%%%%%%%%%%%%%%%%%%%

\subsection{Dissipative and dissipationless contributions}
The results above include both dissipative (also referred to as ohmic) as well as dissipationless (or Hall-like) contributions~\cite{stepan2022on, debottam2024quantum}.  
%After segregating the nonlinear conductivity based on their scattering time dependence, we now decompose them into the dissipative (Ohmic) and dissipationless (Hall) parts by following Ref.~\cite{stepan2022on}. 
As the name suggests, the part of nonlinear conductivity that leads to heat dissipation is called dissipative. In contrast, the remaining part of the conductivity, not participating in heat dissipation is identified as dissipationless or Hall conductivity. The heat dissipated by the nonlinear thermoelectric current is proportional to ${\bm j}\cdot{\bm E}_T \equiv T^2 \alpha_{a;bc} E^a_T E^b_T E^c_T$, where Einstein summation convention over repeated indices is implied. This is analogous to the electric current where the heat dissipated is proportional to ${\bm j}\cdot{\bm E}$, with ${\bm E}$ being the electric field. See Sec.~S3 of the SM~\cite{Note1} for details of the heat dissipation for the thermoelectric current. Using this, the thermoelectric transport coefficients can be segregated into dissipative and dissipationless contributions. 
%By summing up all spatial indices, we find that 
We find that the completely symmetric part of the nonlinear conductivity tensor in all spatial indices contributes to heat dissipation. Thus, for a given third-rank nonlinear conductivity tensor $\alpha_{a;bc}$, we first construct a thermoelectric conductivity tensor that is symmetric in field indices ${\bar \alpha}_{a;bc} = (\alpha_{a;bc} + \alpha_{a;cb})/2$. Then, we compute the Ohmic or dissipative part by symmetrizing $\alpha_{a;bc}$ over all the spatial indices. Using this, we deduce the Hall or dissipationless part by subtracting the Ohmic part from the ${\bar \alpha}_{a;bc}$. We present the resulting nonlinear Ohmic and Hall conductivities in Table~\ref{tab:table1}. A detailed calculation of Ohmic and Hall conductivities is presented in Sec.~4 of the SM~\cite{Note1}.

In Table~\ref{tab:table1}, we categorize the nonlinear thermoelectric current into its dissipative and dissipationless components. Each of these has contributions from three mechanisms: nonlinear Drude, anomalous, and intrinsic. We find that the NLD current is fully dissipative, while the NLA current is entirely dissipationless. It is important to note that the longitudinal (Seebeck) response is always dissipative, whereas the transverse (Nernst) response can be either dissipative or dissipationless. The NLD mechanism contributes to both the Nernst and Seebeck effects, while the NLA mechanism is limited to the Nernst effect. Interestingly, the NLI current encompasses both dissipative and dissipationless contributions. The nonlinear dissipative conductivity can occur in both longitudinal and transverse directions influencing both Seebeck and Nernst effects. In contrast, the dissipationless NLI and NLA conductivity exists only in the transverse direction, contributing exclusively to the Nernst effect. We summarize these findings schematically in Fig.~\ref{fig:HO_schematic}. 

\subsection{Impact of crystalline symmetries}
To understand these responses better, we perform symmetry analysis on the current equation, $j_a = T^2 \alpha_{a;bc}(\tau) E^b_T E^c_T$. Here, the explicit time dependence is encoded in scattering time $\tau$. In the presence of inversion (${\m P}$) symmetry, the electric current and thermal field change sign while the scattering time remains unchanged. Thus, we have $j_a = -j_a$ under inversion symmetry, which forbids any thermoelectric response at the second order in the temperature gradient. However, in the presence of time-reversal (${\m T}$) symmetry, the electric current and scattering time reverse their sign. At the same time, the thermal field remains unaltered, which leads to the following constraint: $\alpha(\tau) = -\alpha(-\tau)$. Thus, time-reversal symmetry forces the thermoelectric conductivity tensors having odd powers of $\tau$ to be finite and those with even power of $\tau$ to vanish. Alternately,  $\alpha^{\rm NLA}_{abc}$ is the only non-vanishing contribution in materials preserving time-reversal symmetry. 
%while $\alpha^{\rm NLD}_{a;bc}$ and $\alpha^{\rm NLI}_{a;bc}$ vanishes. Lastly, 
In the presence of inversion-time reversal (${\m PT}$) symmetry with broken ${\m P}$ and ${\m T}$ symmetries, the Berry curvature becomes zero all over the Brillouin zone, and this forces $\alpha^{\rm NLA}_{a;bc}$ to vanish, while $\alpha^{\rm NLD}_{a;bc}$ and $\alpha^{\rm NLI}_{a;bc}$ remain finite.  
As a consequence, ${\m PT}$ symmetric crystalline materials are a good platform for exploring and probing the nonlinear intrinsic thermoelectric response. 

By analyzing a material's crystalline symmetries, we can predict which response tensors will be finite. Symmetry analysis has previously identified the magnetic point groups that support dissipationless transverse, as well as dissipative longitudinal and transverse, electrical response tensors~\cite{zhnag2023symmetry, stepan2022on}. Since thermoelectric and electrical response tensors follow the same transformation rules, the symmetry analysis in Refs.~[\onlinecite{stepan2022on, zhnag2023symmetry}] also applies to the thermoelectric response tensors in Table~\ref{tab:table1}.

%%%%%%%%%%%%%%%%%%%%%%%%%%%%%%%%%%%%%%%%%%%%%%%%%%%%%%%%%%%%%%%%
\section{Intrinsic Nernst and Seebeck response in $\textbf{CuMnAs}$}

As discussed above, ${\m PT}$ symmetric systems are good candidate materials for studying intrinsic nonlinear thermoelectric responses. To demonstrate this, we use ${\rm CuMnAs}$, which has antiferromagnetic ordering with opposite spins lying on a bipartite lattice. Such spin arrangement locally breaks the ${\m P}$ and ${\m T}$ symmetry. However, the combined ${\m PT}$ symmetry is preserved under exchanging the sublattices (denoted below by $A$ and $B$) along with the flipping of oppositely aligned spins~\cite{wadley2013tetragonal, watanabe2021chiral}. The model Hamiltonian for ${\rm CuMnAs}$ is given by~\cite{smejkal2017electric, watanabe2021chiral}
\be\label{eq:ham}
{\m H}(\kb) = 
\begin{pmatrix}
\epsilon_0(\kb) + {\bm h}_A (\kb)\cdot {\bm \sigma} & V_{AB}(\kb) \\
V_{AB}(\kb) & \epsilon_0(\kb) + {\bm h}_B (\kb)\cdot {\bm \sigma}
\end{pmatrix}~.
\ee 
Here, $V_{AB} = -2\tilde{t} \cos{k_x/2}\cos{k_y/2}$ and $\epsilon_0 (\kb) = -t(\cos{k_x} + \cos{k_y})$, with $t$ and $\tilde{t}$ being the hopping between the orbitals of the same and different sublattices, respectively. In Eq.~\eqref{eq:ham}, we include the sublattice-dependent spin-orbit coupling and the magnetization field in ${\bm h}_B (\kb) = - {\bm h}_A (\kb)$, where ${\bm h}_A (\kb) = \{ h^x_{\rm AFM} - \alpha_R \sin{k_y} + \alpha_D \sin{k_y}, h^y_{\rm AFM} + \alpha_R \sin{k_x} + \alpha_D \sin{k_x}, h^z_{\rm AFM} \}$. Here, $\alpha_R$ ($\alpha_D$) represents the Rasbha (Dresselhaus) spin-orbit coupling, with ${\bm \sigma} = \{ \sigma_x, \sigma_y, \sigma_z\}$ being the Pauli matrices. 
The energy eigenvalues of this Hamiltonian are given by $\e(\kb) = \epsilon_0 \pm \sqrt{V^2_{AB} + h^2_{Ax} + h^2_{Ay} + h^2_{Az}}$, where $+(-)$ denotes conduction(valence) band. The presence of finite $\epsilon_0$ breaks the particle-hole symmetry and $h_{Ax}(-k_x, -k_y) \ne h_{Ax}(k_x, k_y)$ leads to $\e(-\kb) \neq \e(\kb)$.

For the CuMnAs Hamiltonian in Eq.\eqref{eq:ham}, we numerically compute both its band dispersion and quantum metric components. Figure~\ref{fig:cumnas}\textbf{a} shows the 3D band dispersion of CuMnAs in momentum space, which has a finite band gap. In Figures~\ref{fig:cumnas}\textbf{b-d}, we present the phase-space distribution of the quantum metric components ${\m G}^{xx}$, ${\m G}^{xy}$, and ${\m G}^{yy}$ for the conduction band. These quantum metric components are concentrated near the band edges. Notably, the ${\m G}^{xx}$ distribution forms a dipole-like pattern, while ${\m G}^{xy}$ and ${\m G}^{yy}$ exhibit quadrupole characteristics. Utilizing the band structure and quantum metric of the CuMnAs model, we numerically calculate the dissipative and dissipationless components of the nonlinear thermoelectric current.

We present the dependence of the nonlinear thermoelectric components on the chemical potential in CuMnAs in Fig.~\ref{fig:nl_cond}. 
The figure is divided into three panels: the left panel presents the total nonlinear dissipative Seebeck conductivity, the middle panel displays the nonlinear dissipative Nernst conductivity, and the right panel shows the nonlinear dissipationless Nernst conductivity. In each panel, the black dashed line represents the total conductivity, while the blue, red, and orange lines correspond to the NLD, NLA, and NLI contributions, respectively. Notably, there is no NLA contribution in any panel, while the NLI component dominates the thermoelectric response in all panels. The absence of NLA conductivity arises from the ${\m PT}$ symmetry of CuMnAs, which makes the Berry curvature vanish across the Brillouin zone. Additionally, while the NLD current scales quadratically with the scattering time $\tau$, the NLI conductivity remains independent of $\tau$. These findings demonstrate that, for reasonable values of the scattering timescale~\cite{oles2024kramers}, the quantum metric-induced NLI thermoelectric responses dominate over the NLD contribution in ${\rm CuMnAs}$.

\section{Discussion}
In experiments, the measured thermoelectric conductivity reflects the sum of all contributions. To identify the different contributions from the measured data, we propose the following. The longitudinal nonlinear Seebeck conductivity is always dissipative, even though a part of it can arise from an intrinsic mechanism completely independent of the scattering timescale. To identify the intrinsic and Drude contributions, we can make a parametric plot of the measured nonlinear Seebeck conductivity against the measured longitudinal electric conductivity $\sigma_{xx} \propto \tau$ in the linear response regime, by varying some system parameters. In this parametric plot, the contribution independent of $\sigma_{xx}$ can be identified as the intrinsic contribution, while the contribution varying as $\sigma_{xx}^2$ can be identified as the Drude contribution. 

For the nonlinear Nernst response, if the system is nonmagnetic, then the dominant contribution to the second-order nonlinear thermoelectric response is the Berry curvature dipole contribution, as the NLI and NLD contributions vanish in this regime. For inversion-broken magnetic systems, the second-order Nernst effect can have all three contributions. However, the three different contributions have different $\tau$ dependence and can be distinguished via the parametric plot with $\sigma_{xx}$ as discussed above. 
%However, the two contributions can be distinguished by using the fact that the Berry curvature (Drude) contribution is odd (even) under the interchange of the current and the temperature gradient direction. 

We have primarily discussed the longitudinal and transverse thermoelectric responses. However, our results in Table~\ref{tab:table1} are more general and include the mixed thermoelectric responses. 
%\sout{For a two-dimensional system in the $xy$-plane, all possible elements of the nonlinear thermoelectric tensor are ${\bar \alpha}_{x;xx}, ~{\bar \alpha}_{x;xy},~{\bar \alpha}_{x;yx}, ~{\bar \alpha}_{x;yy}, ~{\bar \alpha}_{y;xx},~{\bar \alpha}_{y;xy},~{\bar \alpha}_{y;yx}$ and ${\bar \alpha}_{y;yy}$.} 
In a two-dimensional system confined to the $xy$-plane, the relevant elements of the nonlinear thermoelectric tensor are ${\bar \alpha}_{x;xx}, {\bar \alpha}_{x;xy}, {\bar \alpha}_{x;yy}, {\bar \alpha}_{y;xx}, {\bar \alpha}_{y;xy}$, and ${\bar \alpha}_{y;yy}$. Among these, ${\bar \alpha}_{x;xx}$ and ${\bar \alpha}_{y;yy}$ directly contribute to the nonlinear dissipative Seebeck current, while ${\bar \alpha}_{x;yy}$ and ${\bar \alpha}_{y;xx}$ contribute to both dissipative and dissipationless components of the nonlinear Nernst current. The off-diagonal elements, ${\bar \alpha}_{x;xy} = {\bar \alpha}_{x;yx}$ and ${\bar \alpha}_{y;xy} = {\bar \alpha}_{y;yx}$, represent a mixed response that does not directly correspond to either Seebeck or Nernst currents. However, these mixed elements play a crucial role in light-induced thermoelectric responses, where a focused laser can generate temperature gradients along both the $x$ and $y$ directions simultaneously. This opens exciting possibilities for controlling thermoelectric effects using light for potential device applications.

One experimental technique to explore the thermoelectric responses is to use a finite frequency ($\omega$) current to induce the temperature gradient. In this case, the temperature gradient is proportional to the square of the current and manifests at the frequency $2\omega$. In such experiments, lock-in measurements can identify the second-order nonlinear thermoelectric response by measuring the Seebeck or the Nernst voltage at frequency $4\omega$.

Beyond our present calculations, additional contributions to the  
second-order thermoelectric responses may arise from extrinsic scattering contributions such as the side-jump and skew scattering. Developing a full understanding of these contributions, along with the development of the scaling analysis for identifying different contributions\cite{Du2019disorder, anyuan2023quantum, huang2023scaling}, can be a potential future research project.  

\section{Conclusion}

In this work, we discovered a novel contribution to nonlinear thermoelectric current, the `nonlinear intrinsic current,' which is independent of scattering time and originates from the quantum metric tensor. This previously unexplored contribution arises from interband coherence effects in multi-band quantum systems. Alongside this intrinsic contribution, we also calculated the nonlinear Drude and anomalous thermoelectric currents using the quantum kinetic theory framework.

Our findings reveal that the nonlinear intrinsic thermoelectric conductivity includes both Nernst and Seebeck contributions. While the longitudinal Seebeck contribution is always dissipative, the transverse Nernst contribution can be both dissipative and dissipationless. These contributions can be distinguished by identifying the symmetric and anti-symmetric parts of the conductivity tensor under the exchange of the current and temperature gradient directions. In contrast, the Drude Seebeck contributions are always dissipative, whereas the nonlinear anomalous Nernst effect is always dissipationless.

We demonstrated that in nonmagnetic and non-centrosymmetric systems, only the Berry curvature-induced nonlinear anomalous Nernst effect exists. The intrinsic nonlinear thermoelectric response is finite only in magnetic and non-centrosymmetric systems. In particular, in a ${\m PT}$ symmetric system such as a bipartite antiferromagnet, the nonlinear anomalous Nernst effect vanishes, allowing the intrinsic thermoelectric response to dominate. We illustrated this in a ${\m PT}$ symmetric ${\rm CuMnAs}$ system at low temperatures and argued that the distinct dependence of these current components on $\tau$ provides a tool for identifying them in experiments. Our study highlights the significance of band geometry-induced intrinsic nonlinear Seebeck and Nernst currents and motivates further exploration of nonlinear thermoelectric responses for developing efficient energy harvesting devices.

%%%%%%%%%%%%%%%%%%%%%%%%%%%%%%%%%
\section{Acknowledgements}
H.V. acknowledges the Ministry of
Education, Government of India, for financial support through the Prime Minister’s Research Fellowship.
%%%%%%%%%%%%%%%%%%%%%%%%%%%%%%%%%%
\bibliography{ref_new}
\end{document}